\documentclass[12pt]{iopart}

\usepackage{amssymb,amsfonts}
\usepackage{graphicx}

\newcommand{\beq}{\begin{equation}}
\newcommand{\eeq}{\end{equation}}
\newcommand{\bea}{\begin{eqnarray}}
\newcommand{\eea}{\end{eqnarray}}

\begin{document}

\title[Initial Data of Arbitrary Binaries]{Multi-Domain Spectral Method for Initial Data of Arbitrary Binaries in General Relativity}

\author{Marcus Ansorg}

\address{Max-Planck-Institut f\"ur Gravitationsphysik,
  Albert-Einstein-Institut, Am M\"uhlenberg 1, D-14476 Golm, Germany}
\ead{marcus.ansorg@aei.mpg.de}

\begin{abstract}
We present a multi-domain spectral method to compute initial data of binary systems in General Relativity. By utilizing adapted conformal coordinates, the vacuum region exterior to the gravitational sources is divided up into two subdomains within which the spectral expansion of the field quantities is carried out. If a component of the binary is a neutron star, a further subdomain covering the star's interior is added. As such, the method can be used to construct arbitrary initial data corresponding to binary black holes, binary neutron stars or mixed binaries. In particular, it is possible to describe a black hole component by the puncture ansatz as well as through an excision technique. First examples are given for binary black hole excision data that fulfill the requirements of the quasi-stationary framework, which combines the Conformal Thin Sandwich formulation of the constraint equations with the Isolated Horizon conditions for black holes in quasi-equilibrium. These numerical solutions were obtained to extremely high accuracy with moderate computational effort. Moreover, the method proves to be applicable even when tending toward limiting cases such as large mass ratios of the binary components.
\end{abstract}



\section{Introduction} \label{s:intro}

The past year has seen considerable progress in the numerical calculation of the dynamics of binary black holes. Different groups were able to demonstrate computations describing multiple orbits, the merger as well as the final ring-down phase with the corresponding gravitational radiation emitted \cite{Pretorius2005a, Pretorius2006, Campanelli2006, Baker2006, Campanelli2006a, Baker2006a, Herrmann2006, Baker2006b, Campanelli2006b, Campanelli2006c, Sperhake2006, Bruegmann2006, Gonzalez2006, Scheel2006}. While most of these groups use initial data arising from the puncture ansatz \cite{Brandt1997}, different initial data, which result from the quasi-stationary framework of Conformal Thin Sandwich equations \cite{York1999, Pfeiffer2003} (see also review \cite{Cook2000}) and Isolated Horizon boundary conditions \cite{Cook2002,Cook2004} (see also \cite{Pfeiffer2003a}; for reviews on the Isolated Horizon formalism see \cite {Ashtekar2005,Booth2005,Gourg2006}) have also been chosen. As a consequence, the numerics of dynamical binary black holes has reached a stage in which important astrophysical implications can be addressed, e.g.~the accurate calculation of the final black hole's ``kick'' velocity which appears for unequal-mass binaries due to the asymmetric loss of linear momentum in the gravitational radiation emitted during the merger (see e.g.~\cite{Gonzalez2006}).

The initial data used in these calculations do not represent an exact astrophysical initial situation of distant orbiting black holes, but only an approximation to this. In particular, through a specific (often mathematically convenient) fixing of ``free'' components in the initial data construction, an undesired and uncertain amount of artificial gravitational radiation is contained in the data. In the calculation of the dynamics of neutron stars as well as that of mixed binaries (consisting of a black hole and a neutron star) the same issues arise. It is therefore necessary to check the numerically calculated astrophysical results with respect to a large variety of different types of binary initial data.

Indeed, a large variety is available because of the fact that, in order to construct a specific initial data set, one has to
\begin{enumerate}
	\item choose a formulation of the constraint equations,
	\item fix the free components in the data,
	\item for black holes: choose between the puncture\footnote{In the puncture methods, a special pole-like structure of the singularity (called the ``puncture'') inside the black hole is assumed which can be taken into account by a specific ansatz for the initial data. Therefore the relevant space for the constraint equations is all of $\mathbb{R}^3$.} and the excision method\footnote{The excision techniques solve the constraints only in the exterior of an excised spheroid within which the singularity is located.},
	\item for black holes with excision technique: fix the formulation of the boundary conditions needed at the excision surface.
\end{enumerate}

In this paper we introduce a new multi-domain spectral method to compute arbitrary initial data of binary systems in General Relativity. Apart from the freedom to provide any type of data, we aim at an extremely high numerical accuracy close to machine precision, even when tending toward limiting cases such as large mass ratios or large distances of the binary components. Because of their exponential convergence properties, we utilize pseudo-spectral methods, which also have been used e.g. in \cite{Bonazzola1999a, Gourgoulhon 2001, Gourgoulhon2002, Grandclement2002, Pfeiffer2002, Pfeiffer2003a, Cook2004, Ansorg2004, Ansorg2005, Uryu2005, Pfeiffer2005, Caudill2005, Grandclement2006} for the construction of initial data. 

Conformal coordinate mappings for the construction of initial data in General Relativity have been explored in the literature, see e.g. \cite{Cadez1971, Smarr1976, Cook1991, Bromley2005}. The basic feature of the method described here is a specific conformal coordinate mapping which permits the identification of the entire spatial domain as the image of a small number of compact subdomains, introduced such that the data are analytic there. Consequently, a rapidly converging spectral expansion arises\footnote{Note that the corresponding coordinate curves are in general not smooth at common domain boundaries, see figures \ref{f:Double_Punc} -- \ref{f:two_star}. This does not lead to a loss of accuracy in the numerical scheme since the spectral expansions are performed separately in each subdomain and the communication between them is realized through transition conditions (continuous behaviour of functions and normal derivatives).}.
In particular, the exterior vacuum region is divided up into two subdomains, and for each neutron star component of the binary a subdomain covering all of the star's interior is added. Thus we obtain a two-domain spectral method for binary black holes, a three-domain spectral method for mixed binaries and a four-domain spectral method for neutron star binaries. The method can be used to compute the following types of initial data:
\begin{enumerate}
	\item Binary black hole puncture data
	\item Mixed binary black hole puncture-excision data
	\item Binary black hole excision data
	\item neutron star - black hole binary data; black hole: puncture
	\item neutron star - black hole binary data; black hole: excision
	\item Binary neutron star data
\end{enumerate}

The paper is organized as follows. In \sref{s:conf_map} the conformal mapping in question is introduced. \Sref{s:spec_maps} describes in which manner the partition into subdomains as well as their individual mappings can be undertaken for each one of the above types of initial data. The pseudo-spectral scheme which takes all subdomains into account is explained in \sref{s:pseudo_spectral}. Finally, examples are given in \sref{s:examples} for binary black hole excision data that fulfill the requirements of the above quasi-stationary framework. The corresponding code is of low computational cost, allowing one to do the calculations on a laptop computer. 

Note that we use units in which the speed of light as well as the gravitational constant are equal to one. An overbar denotes complex conjugation.

\section{The conformal coordinate mapping}\label{s:conf_map}

Our starting point is a Cartesian-like coordinate system covering all of the initial $t=\mbox{constant}$ hypersurface.
We denote these coordinates by $(x,y,z)$ defined on the interval $(-\infty,\infty)$. The gravitational sources are assumed to be located at finite coordinate values, thus describing an asymptotically flat spacetime. In particular, we choose the coordinates such that the binary source be aligned along the $x$-axis and the $y$-$z$-plane be located between the binary components. That is, for binary black hole puncture data, we place the punctures at the coordinate values
\beq\label{e:punct_coord}
	x=\pm\, b, \qquad y = 0 = z,\qquad(b>0).
\eeq
In the case of sources that are either black holes, which are given by horizon boundary values at the surface of an excised spheroidal shell, or spheroidal neutron stars, we describe the surface of the spheroidal shells parametrically by
\beq\label{e:spher_coord}
	\begin{array}{lcl}
		x &=& S_\pm(\vartheta,\varphi)\cos\vartheta\pm\, b , \\[3mm ]
		y &=& S_\pm(\vartheta,\varphi)\sin\vartheta\cos\varphi, \\[3mm]
		z &=& S_\pm(\vartheta,\varphi)\sin\vartheta\sin\varphi,	
	\end{array}
\eeq
where the positive smooth functions $S_\pm$ can be expanded with respect to spherical harmonics $Y_l^m(\vartheta,\varphi)$. For the examples considered in this paper we will refer to the simplest case, in which the shells are exactly spherical. However, the method also allows more sophisticated as well as unknown surface functions which, similar to the ``free'' boundary value problems for axisymmetric and stationary fluid configurations, are determined through some quasi-stationary condition imposed on the surface of the shells\footnote{For neutron stars, the vanishing of the energy density at the surface provides such a condition, see \cite{Baumgarte2003} for a review.}.

\begin{figure}
	\centerline{\includegraphics[scale=0.7]{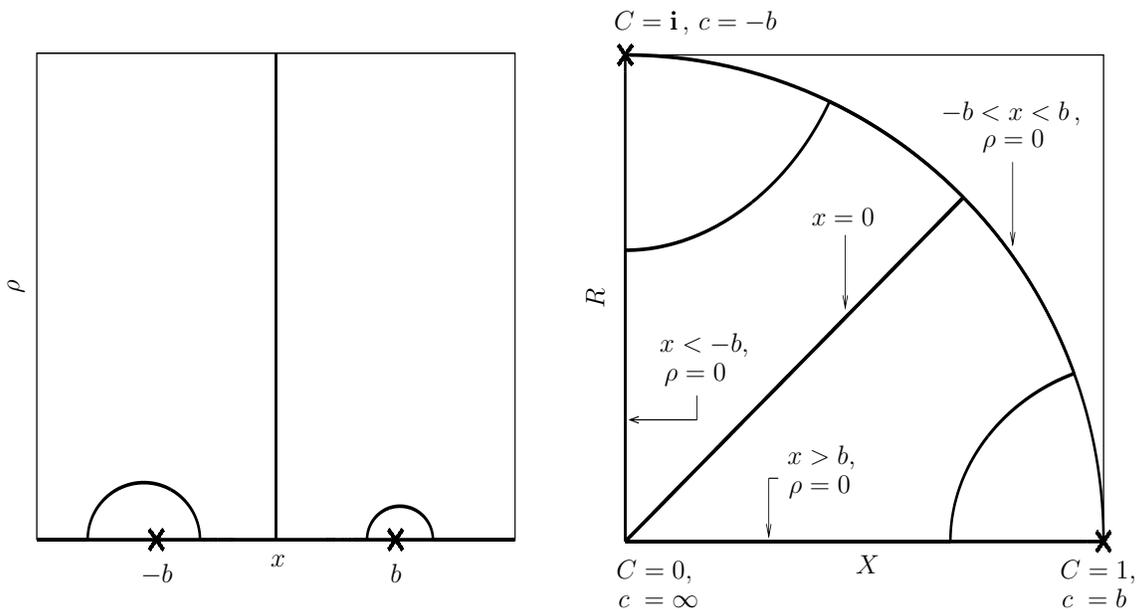}}
	\caption[]{
		\label{f:conf_map}
		Illustration of the conformal mapping $c=\frac{1}{2}b\,(C^2+C^{-2})$ with $c=x-\mbox{\bf i}\, \rho$ and $C=X+\mbox{\bf i}\, R$. Left and right panel show a $(\varphi=\mbox{constant})$-slice in cylindrical coordinates $(x,\rho)$ and the corresponding domain in the coordinates $(X,R)$ respectively. The locations of the punctures are marked with crosses, and the curves around these points indicate spheroidal shells for excised black holes or neutron stars.
	}
\end{figure}

If we are dealing with mixed binaries consisting of a puncture black hole and either a further excised black hole or a neutron star, we place the puncture at one of the points \eref{e:punct_coord} and the shell according to (\ref{e:spher_coord}) on the other side of the $x$-axis. The corresponding setting is illustrated in the left panel of \fref{f:conf_map}.

The introduction of appropriate conformal coordinate mappings has been explored in \cite{Ansorg2004} and \cite{Ansorg2005} in order to calculate binary black hole puncture and excision data respectively. For punctures the key idea was a ``folding'' of the $x$-axis about their locations \eref{e:punct_coord}, in order to obtain coordinates in which the data are analytic there. In a similar fashion, the ``folding of infinity'' in \cite{Ansorg2005} yields binary black hole excision data that are regular at the location of compactified infinity. In this paper we combine the two concepts and propose a conformal coordinate mapping that realizes the following requirements:
\begin{enumerate}
	\item Compactification of spatial infinity,
	\item Regularity of the data at compactified infinity,
	\item Regularity of the data at the points \eref{e:punct_coord}
	(for puncture and neutron star data),
	\item Identification of subdomains within which the data are analytic,
	\item Mapping of these subdomains onto rectangular boxes within which the spectral coordinates 
	\beq\label{spec_coords}
		(A,B,\varphi)\in [0,1]\times[0,1]\times[0,2\pi)
	\eeq
	are defined.
\end{enumerate}
We take cylindrical coordinates $(x,\rho,\varphi)$ about the $x$-axis, i.e.
\beq\label{e:cyl_coords}
	x=x,\qquad y=\rho\cos\varphi,\qquad z=\rho\sin\varphi,
\eeq
in order to form the complex combination
\beq\label{e:c_coord}
	c=x - \rmi \rho.
\eeq
The above requirements are met by the conformal transformation
\beq\label{e:conf_map}
	c=\frac{b}{2}(C^2+C^{-2}),
\eeq
where
\beq\label{e:C_coord}
	C=X + \rmi R.
\eeq
Writing \eref{e:conf_map} explicitely gives
\bea
	\nonumber              x    &=& b \left[\frac{1}{\left(R^2+X^2\right)^2}+1\right] \frac{X^2-R^2}{2}, \\
	\label{e:conf_map_xyz} y    &=& b \left[\frac{1}{\left(R^2+X^2\right)^2}-1\right]  R\,X \cos\varphi, \\
	\nonumber              z    &=& b \left[\frac{1}{\left(R^2+X^2\right)^2}-1\right]  R\,X \sin\varphi. 
\eea
The conformal mapping \eref{e:conf_map} is illustrated in \fref{f:conf_map}. It yields the entire spatial domain as the image of the cross-product of a quarter unit circle (on which $X$ and $R$ are defined) and the interval $[0,2\pi)$ over which the angle $\varphi$ varies, i.e.~
\beq\label{e:Images}
	c:\{(X,R): X\geq 0, R\geq 0, X^2+R^2\leq 1\}\times[0,2\pi)\mapsto \mathbb{R}^3\cup\infty.
\eeq
In particular,
\beq
	\begin{array}{ccccccc}
		C  & = & 0                             & \Rightarrow & c & = &\infty \\
		C  & = & \rmi                          & \Rightarrow & c & = & - b \\
		C  & = & 1                             & \Rightarrow & c & = & + b \\
		C  & = & \frac{1}{2}\sqrt{2}\,(1+\rmi) & \Rightarrow & c & = & 0
	\end{array}
\eeq					
The axes $X=0$ and $R=0$ are mapped to the sections $x<-b$ and $x>b$ of the $x$-axis respectively. The section between the punctures, $-b<x<b$  ($\rho=0$), is the image of the quarter circle curve. For $X=R$ we obtain the $\rho$-axis. \Fref{f:conf_map} illustrates how the mapping folds the $x$-axis about the punctures as well as about infinity. 

For the regularity issues at compactified infinity consider the inverse radius 
\beq
	\frac{1}{r}=\frac{1}{\sqrt{x^2 + \rho^2}} = \frac{1}{|c|} = \frac{2C\bar{C}}{b|1+C^4|},
\eeq 
which is regular at $C = 0$. Likewise, consider the distances
\beq
	r_\pm=\sqrt{(x \mp b)^2 + \rho^2} = |c \mp b| = \frac{b}{2C\bar{C}} \left|1 \mp C^2 \right|^2
\eeq
in the vicinity of the puncture points \eref{e:punct_coord}. They are regular at $C^2=\pm 1$ which ensures the regularity of puncture data there.

In cases with excised black holes or neutron stars, the spheroidal shells are indicated in \fref{f:conf_map} by the curves around the puncture points. Corresponding to the surface functions $S_\pm$, these curves may depend on $\varphi$. The figure provides an intuitive partition of the spatial domain, dividing up the exterior vacuum region into two subdomains and taking for each neutron star component of the binary a subdomain covering all of the star's interior. The specific mappings of the spectral coordinates $(A,B,\varphi)$ onto these subdomains for the various cases is the subject of \sref{s:spec_maps}.

\section{Spectral mappings for binary systems}\label{s:spec_maps}

\subsection{A single domain for binary black hole punctures}
\label{s:single_punc}

\begin{figure}
	\centerline{\includegraphics[scale=0.7]{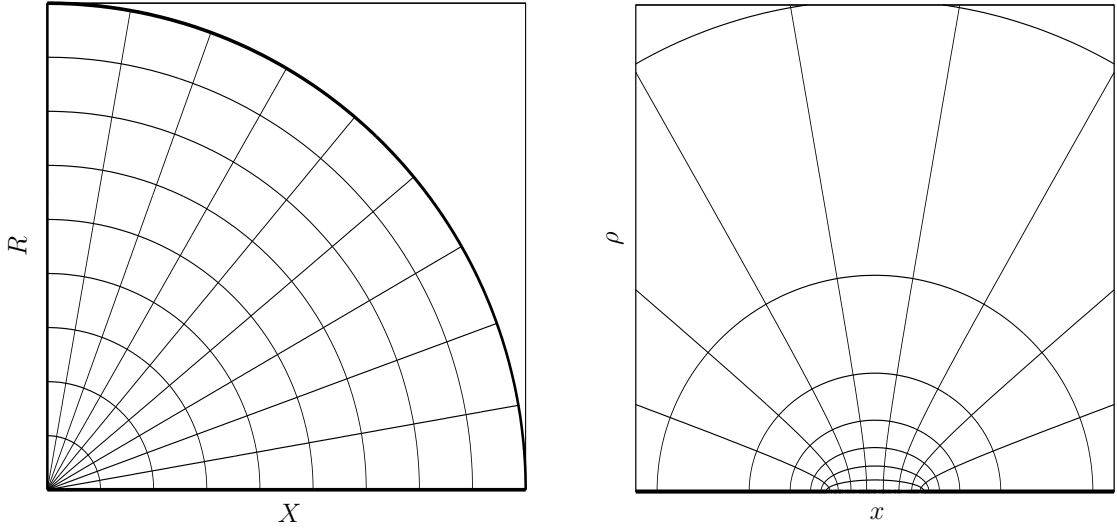}}
	\caption[]{
		\label{f:Single_Punc}  
			Illustration of the spectral mapping \eref{C_of_zeta}-\eref{xieta_of_AB} for single domain binary black hole punctures. The circles $X^2+R^2=\mbox{constant}$ correspond to constant $A$-coordinate lines ($X^2+R^2=\rme^\xi=(1-A)/(1+A)$) and the rays $X/R=\mbox{constant}$ are described by constant $B$-values ($X/R=\cot(\eta/2) = 1/B-1$). The right panel shows the corresponding curves in the cylindrical coordinates $(x,\rho)$.
	}
\end{figure}
For binary black hole puncture data, the quarter unit circle can be covered with a single coordinate patch using polar coordinates, see \fref{f:Single_Punc}. In  fact, with the conformal transformation
\beq\label{C_of_zeta}
	C = \rme^{\zeta/2},
\eeq
where
\beq\label{ReIm_of_zeta}
	\zeta = \xi +\rmi\eta,
\eeq
we obtain
\beq\label{c_of_zeta}
	c=b\,\cosh\zeta,
\eeq
i.e.~ we recover the mapping used in \cite{Ansorg2004} for the calculation of binary puncture data in a single spatial domain. Note that here $\xi\leq 0$ whereas $\xi\geq 0$ in \cite{Ansorg2004}, according to the fact that in that paper the {\em exterior} of a half unit circle has been mapped onto $\mathbb{R}^3$. 

Hence we may write the coordinates $\xi$ and $\eta$ in terms of spectral coordinates in a similar fashion:
\beq\label{xieta_of_AB}
	\begin{array}{lcl}
		\xi &=&-2\mbox{artanh}\,A\,,	\\[3mm]
		\eta&=&\frac{\pi}{2}+2\arctan(2B-1)\qquad(A,B\in[0,1]) \,.
	\end{array}
\eeq
The single domain spectral method for binary black hole puncture data has been used by several groups for the dynamical evolution of binary black hole space-times. For comparable masses and moderate distances of the black holes, this method yields a good convergence rate of the solution as one increases the numerical resolution.

\subsection{Two domains for binary black hole punctures}
\label{s:double_punc}

\begin{figure}
	\centerline{\includegraphics[scale=0.7]{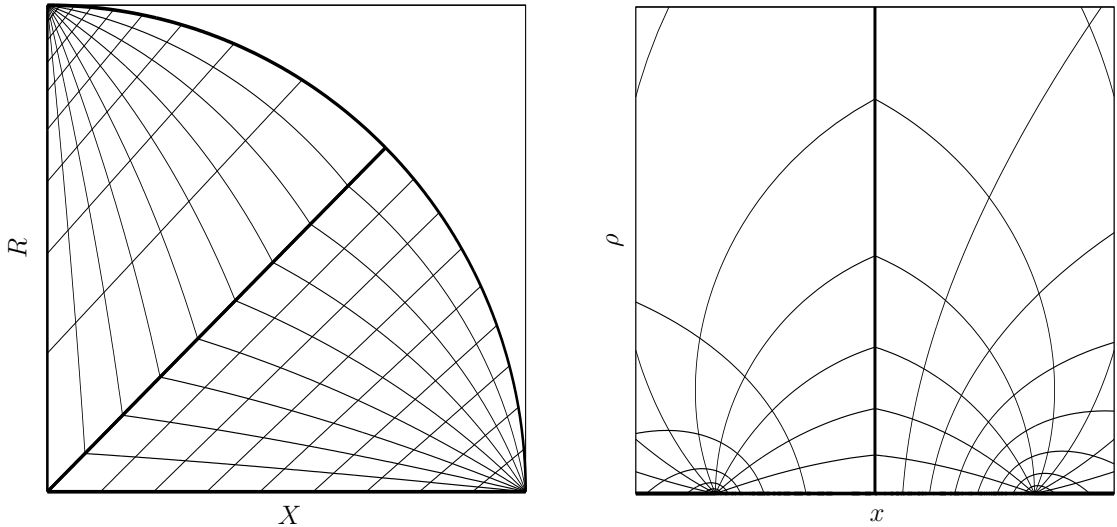}}
	\caption[]{
		\label{f:Double_Punc} 
		Illustration of the spectral mapping \eref{doub_punc_m}-\eref{rescaled_A} for double domain binary black hole punctures. In the vicinity of the punctures, the coordinates $A$ and $B$ resemble radial and angular coordinates respectively. The corresponding curves of constant $A$- and $B$-values are shown in the left and right panel. Specifically, in this example the coordinate lines $A=\mbox{constant}$ form a dense mesh in the vicinity of the puncture $x=-b$, in order to resolve steep gradients. 
	}
\end{figure}

If one considers large distances and/or large mass ratios of the black holes, the gradients of the gravitational field quantites grow large in the vicinity of the punctures. As a consequence, the spectral convergence rate of the single domain method drops significantly. In order to retain the convergence rate in these cases a coordinate mapping can be introduced where one of the coordinates resembles a measure of the coordinate distance from the puncture within the high gradient regime. The rate by which the distance scales with this coordinate can be adjusted such that an appropriate resolution of the steep gradients is achieved. In this manner a double domain method for binary black hole punctures arises. 

A possible transformation which maps the spectral coordinates $(A,B)\in[0,1]^2$ onto the upper vacuum section of the quarter unit circle is given by (see \fref{f:Double_Punc})
\bea\label{doub_punc_m}
	C=\rmi\left[B\exp\left(-\frac{\rmi\pi}{4} A_-\right)+(1-B)(1-A_-)\right].
\eea
Likewise, the lower section is obtained by
\bea\label{doub_punc_p}
	C=B\exp\left(\frac{\rmi\pi}{4} A_+\right)+(1-B)(1-A_+).
\eea
In these formulas the rescaled spectral coordinates $A_\pm$ appear which provide the desired scaling of the coordinate $A$ with the coordinate distance from the puncture. They depend on $b$, the half coordinate distance of the punctures, and some quantity $m_\pm$ that represents a local mass (e.g.~the puncture's bare mass \cite{Brandt1997} or an appropriate horizon mass \cite{Ashtekar2001,Ashtekar2002})\footnote{Here the subscripts $\pm$ refers to the puncture at the coordinate location $x=\pm b\, (\rho=0)$.}:
\beq\label{rescaled_A}
	A_\pm=\frac{\sinh(A\log\epsilon_\pm)}{\sinh(\log\epsilon_\pm)}, \qquad \epsilon_\pm=\frac{m_\pm}{b}.
\eeq
Similar logarithmic rescalings to resolve steep gradients have been explored in \cite{Ansorg2005a}.
\subsection{Two domains for excised binary black holes}
\label{s:exc_bh}

\begin{figure}
	\centerline{\includegraphics[scale=0.7]{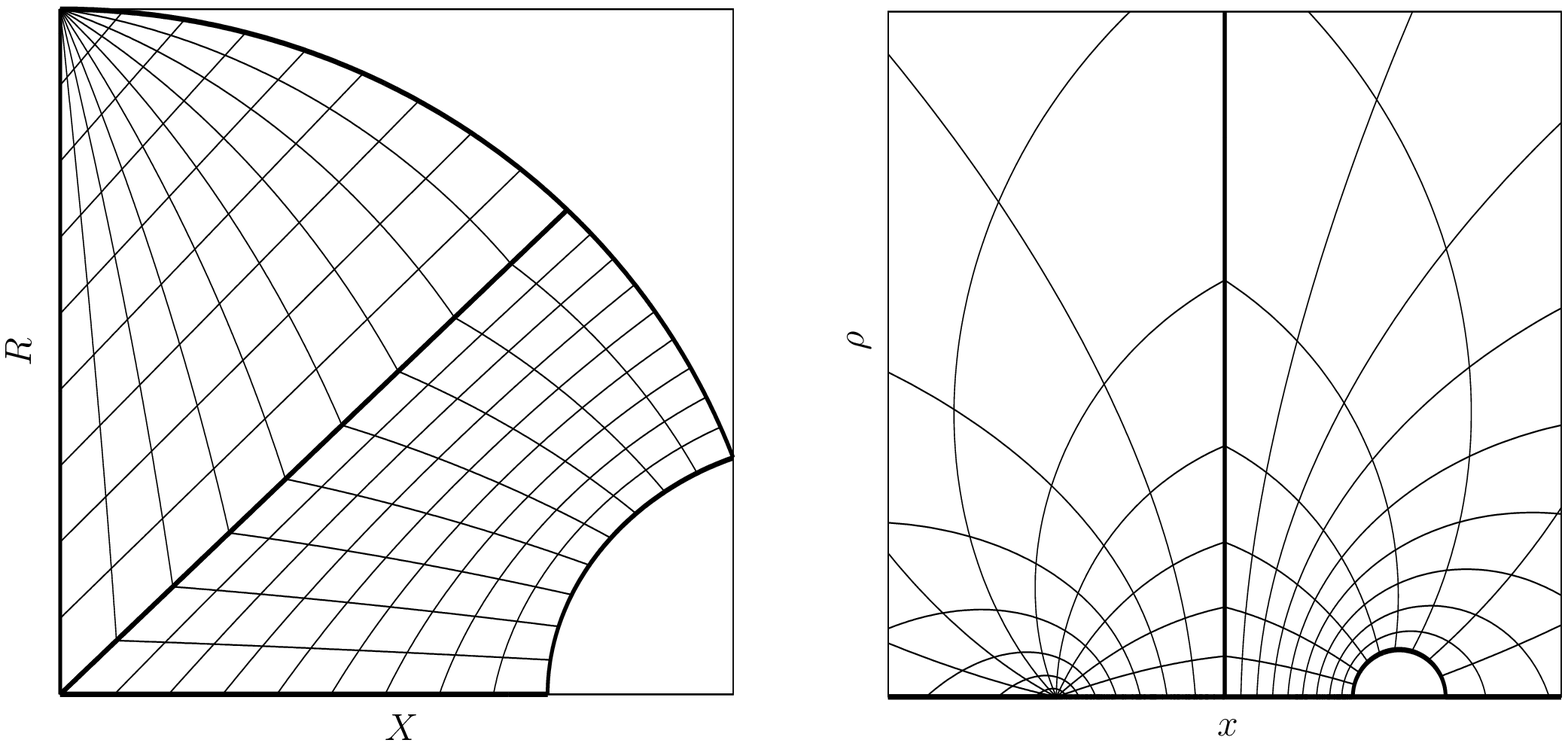}}
	\caption[]{
		\label{f:punc_exc}
		Illustration of the spectral mapping for double domain mixed binary black hole puncture-excision data. The mapping is performed according to the transformations \eref{doub_punc_m} and \eref{exc} in the upper and lower part of the quarter unit circle respectively. In this example, the excision surface is a sphere. Left and right panel display the coordinate curves of constant $A$- and $B$-values.
	}
\end{figure}

\begin{figure}
	\centerline{\includegraphics[scale=0.7]{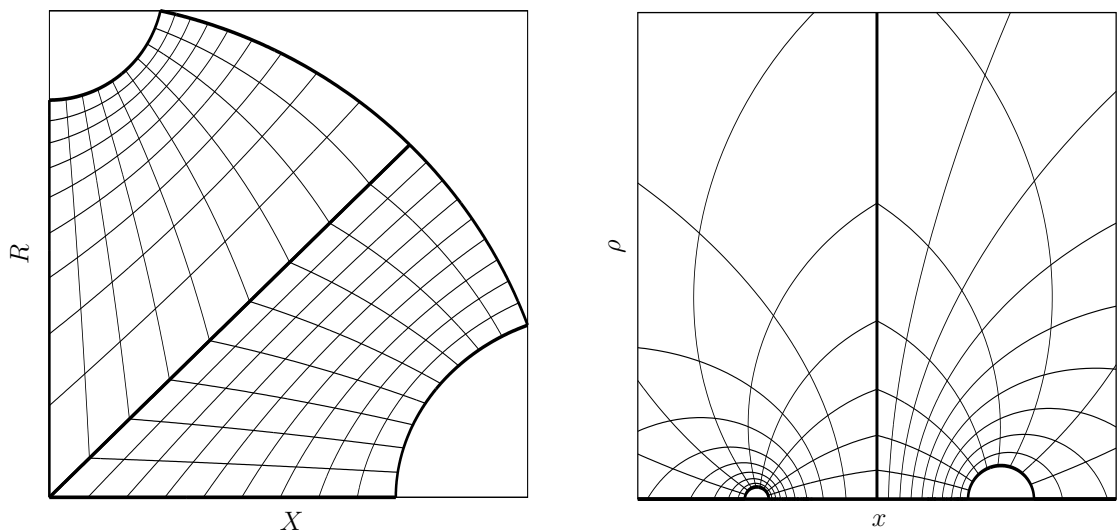}}
	\caption[]{
		\label{f:double_exc}  
		Illustration of the spectral mapping for double domain binary black hole excision data. The mapping is performed according to the transformations \eref{exc} in the upper and lower parts of the quarter unit circle. In this example, the excision surfaces are spheres. Left and right panel display the coordinate curves of constant $A$- and $B$-values.
	}
\end{figure}
In the case of black hole excision data, the transformations \eref{doub_punc_m}, \eref{doub_punc_p} need to be modified in order to take the spheroidal excision surfaces into account. As in \cite{Ansorg2005} we utilize bispherical coordinates \cite{Arfken1970, Moon1988}
\beq\label{bisph_coord}
	\tilde{\zeta}=4\,\mathrm{artanh}\left(C^2\right),
\eeq
in which 
\[
	\Re(\tilde{\zeta})=\mbox{constant}\neq 0,\qquad 0\leq\Im(\tilde{\zeta})\leq\pi
\]
corresponds to exact spherical surfaces about the points \eref{e:punct_coord}. Allowing general shapes, we may describe the surfaces by
\beq\label{surface}
	C^2_\pm(B,\varphi) = \tanh\left[\frac{1}{4}\Big(\sigma_\pm(B,\varphi) + \rmi \pi B\Big)\right],\quad\pm\sigma_\pm(B,\varphi)>0,
\eeq
where the functions $\sigma_\pm$ are related to $S_\pm$ introduced in \eref{e:spher_coord} and take care of the excised spheroidal surface shape, which either can be prescribed or which may be unknown and determined through a quasi-stationary condition to be imposed.

In terms of these functions the mapping of the upper and lower vacuum sections of the quarter unit circle can be written in the following form:
\bea
	C &=& \left(1-A_\pm\right)
		\big[C_\pm(B,\varphi)-B\;C_\pm(1,\varphi)\big] \nonumber\\
		&&+B\exp\left[\rmi\left(\frac{\pi}{4}A_\pm+\left(1-A_\pm\right)\arg C_\pm(1,\varphi)\right)\right].\label{exc}
\eea
Again the rescaled spectral coordinates $A_\pm$ appear to render steep potential gradients, see \eref{rescaled_A}.

The puncture mappings \eref{doub_punc_m}, \eref{doub_punc_p} can be used together with the excision mappings \eref{surface}, \eref{exc}, in order to calculate mixed binary black hole puncture-excision data, see \fref{f:punc_exc}. Double excision black hole binaries are obtained by mapping both subdomains according to \eref{surface} and \eref{exc}, see \fref{f:double_exc}.

\subsection{Domains for neutron stars}
\label{s:stars}

\begin{figure}
	\centerline{\includegraphics[scale=0.7]{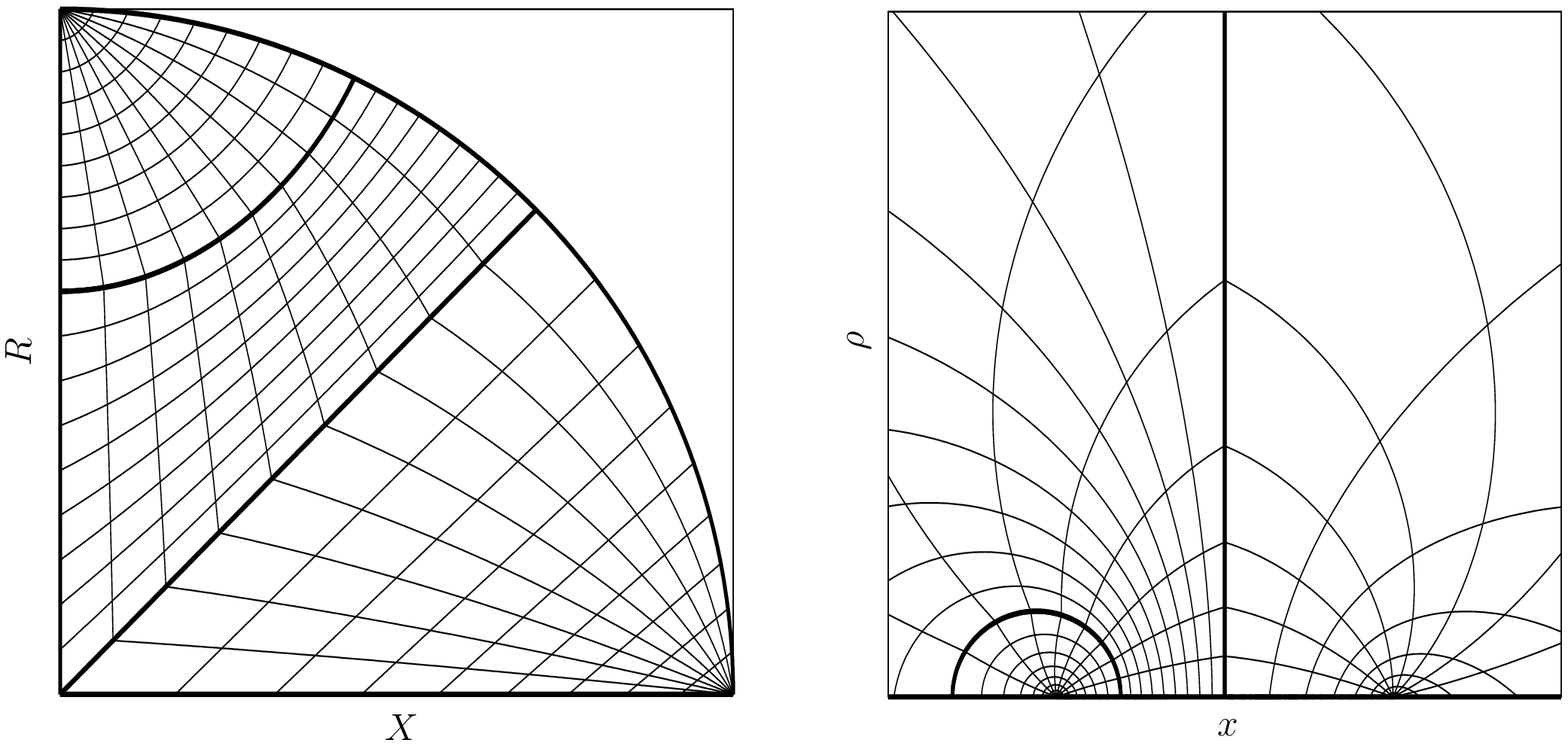}}
	\caption[]{
		\label{f:punc_star}  
		Illustration of the three-domain spectral mapping for mixed binary black hole-neutron star data, where the black hole is described by the puncture method. The mapping is performed according to the transformations \eref{star_m}, \eref{exc} and \eref{doub_punc_p} in the upper and lower part of the quarter unit circle respectively. In this example, the star surface is a sphere. Left and right panel display the coordinate curves of constant $A$- and $B$-values.
	}
\end{figure}

\begin{figure}
	\centerline{\includegraphics[scale=0.7]{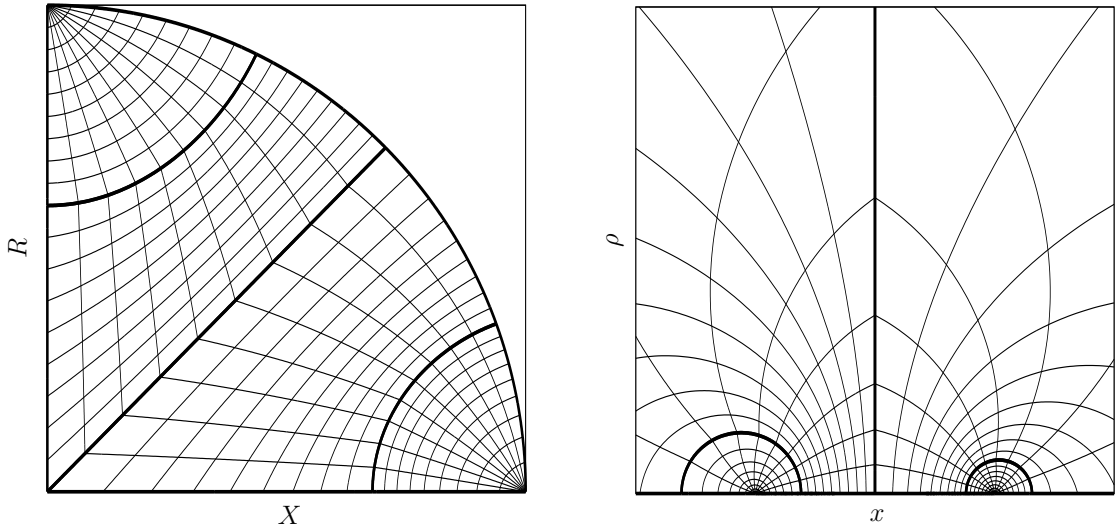}}
	\caption[]{
		\label{f:two_star}
		Illustration of the four-domain spectral mapping for binary neutron star data. The mapping is performed according to the transformations \eref{star_m}, \eref{exc} and  \eref{star_p}, \eref{exc} in the upper and lower part of the quarter unit circle respectively. In this example, the star surfaces are spheres. Left and right panel display the coordinate curves of constant $A$- and $B$-values.
	}
\end{figure}

For the calculation of neutron star data we may take \eref{exc} in order to describe the surrounding exterior vacuum domain. The inner region adds another subdomain. In analogy to expression \eref{exc}, we find a corresponding mapping, through which inner regions (with the same surface shape \eref{surface}) are written in terms of the spectral coordinates:
\begin{figure}
	\centerline{\includegraphics[scale=0.7]{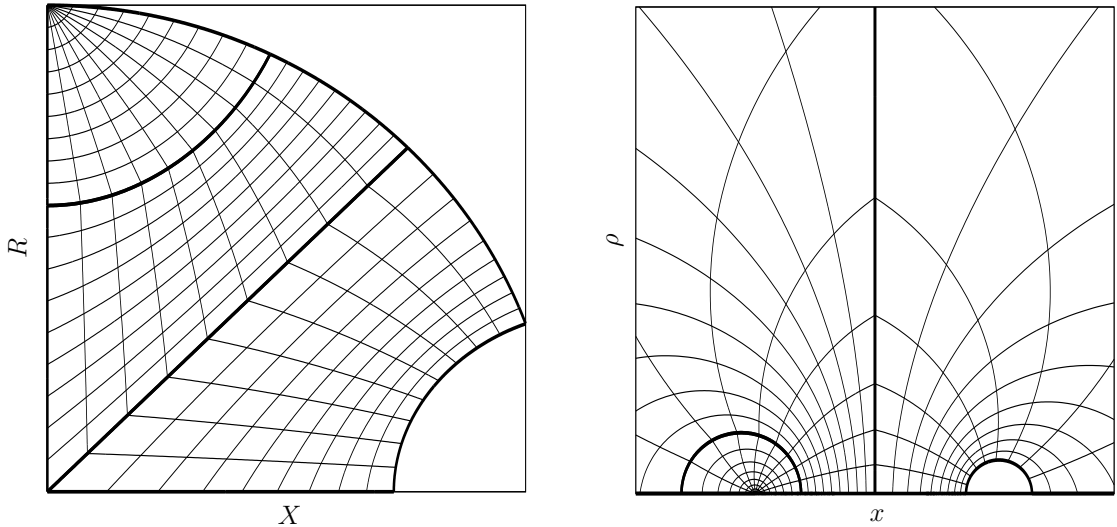}}
	\caption[]{
		\label{f:exc_star}
		Illustration of the three-domain spectral mapping for mixed binary black hole-neutron star data, where the black hole is described by the excision method. The mapping is performed according to the transformations \eref{star_m} and \eref{exc} in the upper and lower parts of the quarter unit circle. In this example, the star and excision surfaces are spheres. Left and right panel display the coordinate curves of constant $A$- and $B$-values.
	}
\end{figure}

\bea
	C &=& \left(1-A_-\right)\big[C_-(B,\varphi)-B\;C_-(1,\varphi)\big] \nonumber\\
		&&+B\exp\left[\rmi\left(\frac{\pi}{2}A_-+\left(1-A_-\right)\arg C_-(1,\varphi)\right)\right]+\rmi\,A_-\left(1-B\right)\label{star_m}
\eea
and
\bea
	C &=& \left(1-A_+\right)\big[C_+(B,\varphi)-B\;C_+(1,\varphi)\big] \nonumber\\*
		&&+B\exp\left[\rmi\left(1-A_+\right)\arg C_+(1,\varphi) \right]+A_+\left(1-B\right)\label{star_p}
\eea
for stars at the negative and positive sections of the $x$-axis respectively. 
In figures \ref{f:punc_star} and \ref{f:exc_star} the mapping for mixed binary data, consisting of a black hole and a neutron star, is illustrated; \fref{f:two_star} displays it for binary neutron stars.

\section{The numerical scheme}
\label{s:pseudo_spectral}

The numerical scheme for calculating binary initial data in the various cases generalizes the methods presented in \cite{Ansorg2005} for binary black hole excision data. Most of the techniques are similar, and we refer to that paper for the details. Differences arise in the case of neutron star data where three or four domains are involved. Moreover, conditions with respect to the surface shapes as well as certain required physical parameter relations need to be taken into account. 

In our spectral approximation we write all potentials, that are involved in the elliptic boundary value problem, in terms of Chebyshev expansions with respect to $A$ and $B$ and Fourier expansions with respect to $\varphi$. For the spectral gridpoints we choose the extrema of the Chebyshev polynomials so as to have gridpoints lying on the boundaries:
\bea
	\nonumber 	A_j&=&\sin^2\left[\frac{\pi j}{2(n_A-1)}\right] , \\
	\label{Gridpoints}
				B_k&=&\sin^2\left[\frac{\pi k}{2(n_B-1)}\right] , \\[2mm]
	\nonumber 	\varphi_l&=&\frac{2\pi l}{n_\varphi},
\eea
where 
\beq
	\label{jkl} 0\leq j<n_A,\quad 0\leq k<n_B,\quad 0\leq l<n_\varphi\,.
\eeq 
The numbers $n_A, n_B$ and $n_\varphi$ describe the spectral expansion orders of our scheme. Here $n_A$ may be different in the various subdomains, in order to allow higher resolution in regimes with steep gradients. However, $n_B$ and $n_\varphi$ are equal in all subdomains, in order to have the same spatial grid points on both sides of the domain boundaries.

The potential values at the above gridpoints are combined in a vector describing the spectral approximation of the desired solution. Moreover, we include in this vector the surface shape values $\sigma_\pm(B_k,\varphi_l)$, if they are to be determined. Likewise, we add certain parameters of the solution that are constrained through specific parameter relations. As an example of this, we may take the orbital angular velocity $\Omega$ of the binary system as a further unknown and determine it through the equality of Komar- and ADM-mass, a condition that arises within a quasi-stationary framework.

Having set up this vector, the collection of 
\begin{enumerate}
	\item elliptic equations valid within the subdomains, 
	\item transition conditions to be imposed at common borders of the subdomains,
	\item specific regularity conditions to be fulfilled at the $x$-axis,
	\item specific regularity conditions at $C^2=\pm 1$ (for puncture and neutron star data),
	\item specific asymptotic boundary conditions at $C=0$ (i.e.~at spatial infinity),
	\item excision surface boundary conditions (for black hole excision data),
	\item surface shape conditions (for unknown, quasi-stationary black hole or neutron star boundaries),
	\item parameter relations (for determining specific physical parameters of the solution),
\end{enumerate}
yields a discrete non-linear system\footnote{The equations and conditions are considered at the discrete gridpoints \eref{Gridpoints}.} which is solved by Newton-Raphson iterations. As in \cite{Ansorg2004, Ansorg2005} the linear step inside this solver is performed with the preconditioned `Biconjugate Gradient Stabilized (Bi-CGSTAB)' method \cite{Barrett93}. A good convergence of this method requires a so-called preconditioning, which we construct in complete analogy to \cite{Ansorg2004, Ansorg2005} through a second order finite difference representation of the Jacobian matrix of the non-linear system. In particular, the preconditioner consists of successive plane relaxations with respect to the planes $\varphi=\mathrm{constant}=\varphi_l$. 

For the convergence of the Newton-Raphson scheme a nearby initial `guess' is essential. It can be taken from previously calculated numerical as well as from known analytical solutions. Typically, only a moderate number of iterations within both the Newton-Raphson and the Bi-CGSTAB methods are needed to calculate the desired finite spectral approximation of the solution.

\section{First examples on binary black hole excision data}
\label{s:examples}

\begin{figure}[t]
	\centerline{\includegraphics[scale=1.05]{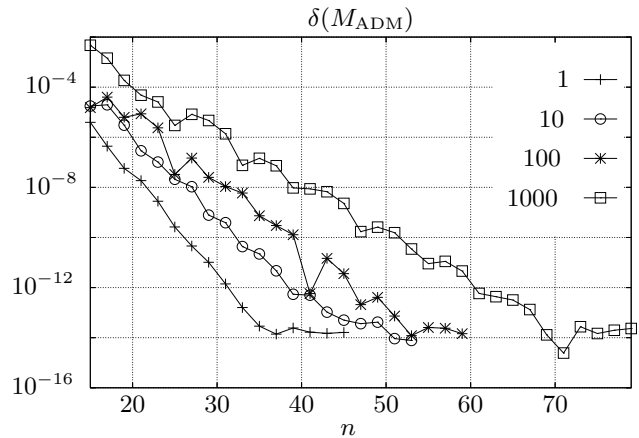}}
	\caption[]{
		\label{f:ADM_axi}
		The convergence of the ADM-mass corresponding to binary black hole initial data sets with vanishing orbital angular velocity $\Omega=0$. The geometrical parameters of the individual configurations are given by $b=5\varrho_-$ and 	$\varrho_+/\varrho_-\in\{1;10;100;1000\}$ where $\varrho_\pm$ describes the coordinate radius of the excised spherical shells. For these axisymmetric calculations $n_A=n_B=n$ and $n_\varphi=3$ have been chosen (in both subdomains). We compared the corresponding results for the ADM-masses to those of reference solutions with $n\in\{50;60;70;80\}$ for the above choices of $\varrho_+/\varrho_-$.
	}
\end{figure}

The representative calculations of binary black hole excision data in \cite{Ansorg2005} are specific examples of the general methods presented in this paper. These initial data fulfill the requirements of the quasi-stationary framework, which combines the Conformal Thin-Sandwich Decomposition with the Isolated Horizon framework.

In the Conformal Thin-Sandwich Decomposition the evolution of the metric between two neighbouring slices $t=\mathrm{constant}$ is considered. For quasi-equilibrium binary data, the constraint and evolution equations of General relativity are treated in a corotating frame of reference in which the time derivatives of the metric quantities are assumed to be small initially. The corresponding Hamiltonian and momentum constraints are elliptic equations to determine the initial data, i.e.~the 3-metric and the extrinsic curvature of the slice. An additional elliptic equation follows from prescribing some value for the time derivative of the trace of the extrinsic curvature. It can be used to provide a specific initial time-slicing of the data.

This quasi-stationary formulation is completed by a set of boundary conditions that control the data at the excision boundaries and at infinity. The excision boundary conditions are given through the Isolated Horizon framework and describe black holes in a quasi-equilibrium state, i.~e.~as instantaneous non-expanding horizons. In particular the following is required:
\begin{enumerate}
	\item 
		Within the initial slice, each excision boundary is an apparent horizon, i.e.~a two-dimensional hypersurface with $S^2$ topology and the property that the outgoing null vectors $k$ possess vanishing expansion.
	\item
		Initially, the apparent horizon is tracked along $k$ and its coordinate location does not move in the time evolution of 	the data.
\end{enumerate}

The examples presented in this paper correspond to so-called conformally flat and maximal Conformal Thin Sandwich data sets of binary corotational black holes that obey the requirements of the Isolated Horizon framework. These data have been previously calculated in \cite{Cook2004, Caudill2005}, see also \cite{Pfeiffer2003a} for further details. 

An ingredient that enters this formulation is the orbital angular velocity $\Omega$ of the binary system. At first we consider $\Omega=0$ which results in specific generalized Misner-Lindquist initial data (they do not represent two orbiting black holes in a quasi-stationary state). As an indicator for the accuracy achieved we monitor the convergence rate of the ADM-mass $M_\mathrm{ADM}$. We find it to be geometric, as exhibited by a roughly linear decrease of the error in \fref{f:ADM_axi}. Note that almost machine accuracy is reached for all ratios considered, thus proving that the method is well suited to the case of extreme mass ratios (and likewise to situations in which the coordinate distance parameter $b$ is extremely large).

\begin{figure}[t]
	\centerline{\includegraphics[scale=1.]{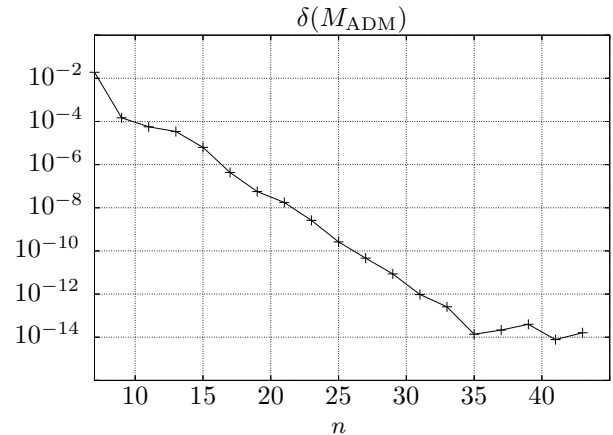}}
	\caption[]{
		\label{f:ADM_nonaxi}
		The convergence of the ADM-mass corresponding to a corotating binary black hole initial data set in a quasi-stationary orbit. The geometrical parameters of the configuration are given by $b=5\varrho_-$ and $\varrho_+=\varrho_-$ where $\varrho_\pm$ describes the coordinate radius of the excised spherical shells. The data are characterized by the equality	$M_{\mathrm{ADM}}=M_{\mathrm{K}}$ through which an orbital angular velocity of $\Omega\approx0.036975/\varrho_-$ emerges. For these calculations the spectral resolutions $n_A=n_B=2n_\varphi+1=n$ have been chosen (in both subdomains). We compared the corresponding results for the ADM-mass to that of the reference solution with $n=51$.
	}
\end{figure}

The second example corresponds to initial data of two corotational black holes in a quasi-stationary orbit. The orbital angular velocity $\Omega$ is obtained by requiring the equality of the ADM-mass $M_\mathrm{ADM}$ and the Komar mass $M_\mathrm{K}$ which can be defined by a suitable surface integral at infinity (see \cite{Gourgoulhon2002, Grandclement2002, Cook2002, Cook2004, Caudill2005}). Again a geometric convergence rate of the resulting ADM-mass arises, see \fref{f:ADM_nonaxi}.

\ack The author wishes to thank J.~L.~Jaramillo and D.~Petroff for numerous valuable discussions.

\section*{References}

\bibliographystyle{unsrt}
\bibliography{Reflink}

\begin{thebibliography}{10}

\bibitem{Pretorius2005a}
Pretorius F 2005
\newblock {\em Phys. Rev. Lett.} {\bf 95} 121101

\bibitem{Pretorius2006}
Pretorius F 2006
\newblock {\em Class.Quant.Grav.} {\bf 23} S529-S552

\bibitem{Campanelli2006}
Campanelli M, Lousto C O, Marronetti P and Zlochower Y 2006
\newblock {\em Phys. Rev. Lett.} {\bf 96} 111101

\bibitem{Baker2006}
Baker J G, Centrella J, Choi D-I, Koppitz M and van Meter J 2006
\newblock {\em Phys. Rev. Lett.} {\bf 96} 111102

\bibitem{Campanelli2006a}
Campanelli M, Lousto C O and Zlochower Y 2006
\newblock {\em Phys. Rev. D} {\bf 73} 061501(R)

\bibitem{Baker2006a}
Baker J G, Centrella J, Choi D-I, Koppitz M and van Meter J 2006
\newblock {\em Phys. Rev. D} {\bf 73} 104002

\bibitem{Herrmann2006}
Herrmann F, Shoemaker D and  Laguna P 2006
\newblock{\em gr-qc/0601026}

\bibitem{Baker2006b}
Baker J G, Centrella J, Choi D-I, Koppitz M, van Meter J and Miller M C 2006
\newblock{\em gr-qc/0603204}

\bibitem{Campanelli2006b}
Campanelli M, Lousto C O and Zlochower Y 2006
\newblock {\em Phys. Rev. D} {\bf 74} 041501

\bibitem{Campanelli2006c}
Campanelli M, Lousto C O, Marronetti P and Zlochower Y 2006
\newblock {\em Phys. Rev. D} {\bf 74} 084023

\bibitem{Sperhake2006}
Sperhake U 2006
\newblock{\em gr-qc/0606079}

\bibitem{Bruegmann2006}
Br{\"u}gmann B, Gonzalez J A, Hannam M, Husa S, Sperhake U and Tichy W 2006
\newblock{\em gr-qc/0610128}

\bibitem{Gonzalez2006}
Gonzalez J A, Sperhake U, Br{\"u}gmann B, Hannam M and Husa S 2006
\newblock{\em gr-qc/0610154}

\bibitem{Scheel2006}
Scheel M A, Pfeiffer H P, Lindblom L, Kidder L E, Rinne O and Teukolsky S A 2006
\newblock {\em Phys. Rev. D} {\bf 74} 104006

\bibitem{Brandt1997}
Brandt S and Br{\"u}gmann B 1997
\newblock {\em Phys. Rev. Lett.} {\bf 78} 3606

\bibitem{York1999}
York J W 1999
\newblock {\em Phys. Rev. Lett.} {\bf 82} 1350

\bibitem{Pfeiffer2003}
Pfeiffer H P and York J W 2003
\newblock {\em Phys. Rev. D} {\bf 67} 044022

\bibitem{Cook2000}
Cook G B 2000
\newblock{\em Living Rev. Relativity} lrr-2000-5 \\
\newblock{\tt http://relativity.livingreviews.org/Articles/lrr-2000-5/}


\bibitem{Cook2002}
Cook G B 2002
\newblock {\em Phys. Rev. D} {\bf 65} 084003

\bibitem{Cook2004}
Cook G B and Pfeiffer H P 2004
\newblock {\em Phys. Rev. D} {\bf 70} 104016

\bibitem{Pfeiffer2003a} 
Pfeiffer H P 2003
\newblock{\em Ph.D. thesis} Cornell University; gr-qc/0510016

\bibitem{Ashtekar2005}
Ashtekar A and Krishnan B 2005
\newblock{\em Living Rev. Relativity} lrr-2004-10  \\
\newblock{\tt http://relativity.livingreviews.org/Articles/lrr-2004-10/}

\bibitem{Booth2005}
Booth I 2005
\newblock {\em Can. J. Phys.} {\bf 83} 1073

\bibitem{Gourg2006} 
Gourgoulhon E and Jaramillo J L 2006
\newblock {\em Phys. Rept.} {\bf 423} 159

\bibitem{Bonazzola1999a}
Bonazzola S, Gourgoulhon E and Marck J A 1999
\newblock {\em Phys. Rev. Lett.} {\bf 82} 892

\bibitem{Gourgoulhon 2001}
Gourgoulhon E, Grandcl\'{e}ment P, Taniguchi K, Marck J-A and Bonazzola S 2001
\newblock {\em Phys. Rev. D} {\bf 63} 064029

\bibitem{Gourgoulhon2002}
Gourgoulhon E, Grandcl\'{e}ment P and Bonazzola S 2002
\newblock {\em Phys. Rev. D} {\bf 65} 044020

\bibitem{Grandclement2002}
Grandcl\'{e}ment P, Gourgoulhon E and Bonazzola S 2002
\newblock {\em Phys. Rev. D} {\bf 65} 044021

\bibitem{Pfeiffer2002}
Pfeiffer H P, Kidder L E, Scheel M A, Teukolsky S A 2002
\newblock {\em Comput. Phys. Commun.} {\bf 152} 253

\bibitem{Ansorg2004}
Ansorg M, Br{\"u}gmann B and Tichy W 2004
\newblock {\em Phys. Rev. D} {\bf 70} 064011

\bibitem{Ansorg2005}
Ansorg M 2005
\newblock {\em Phys. Rev. D} {\bf 72} 024018

\bibitem{Uryu2005}
Uryu K, Limousin F, Friedman J L, Gourgoulhon E and Shibata M 2005
\newblock {\em Phys. Rev. Lett.} {\bf 97} 171101

\bibitem{Pfeiffer2005}
Pfeiffer H P and York J W 2005
\newblock {\em Phys. Rev. Lett.} {\bf 97} 091101

\bibitem{Caudill2005}
Caudill M, Cook G. B., Grigsby J. D. and Pfeiffer H. P. 2006
\newblock{\em Phys.Rev. D} {\bf 74} 064011

\bibitem{Grandclement2006}
Grandcl\'{e}ment P 2006
\newblock {\em Phys. Rev. D} {\bf 74} 124002

\bibitem{Cadez1971} 
{\v C}ade{\v z}  A 1971 
\newblock{\em Ph.D. thesis} University of North Carolina at Chapel Hill

\bibitem{Smarr1976} 
Smarr L, {\v C}ade{\v z}  A, De Witt B and Eppley K 1976 
\newblock {\em Phys. Rev. D} {\bf 14} 2443

\bibitem{Cook1991}
Cook G B 1991
\newblock {\em Phys. Rev. D} {\bf 44} 2983

\bibitem{Bromley2005} 
Bromley B, Owen R. and Price R. H. 2005
\newblock {\em Phys. Rev. D} {\bf 71} 104017

\bibitem{Baumgarte2003}
Baumgarte T W and Shapiro S L 2003
\newblock {\em Phys. Rep.} {\bf 376} 41

\bibitem{Ashtekar2001}
Ashtekar A, Beetle C and Lewandowski J 2001
\newblock {\em Phys. Rev. D} {\bf 64(4)} 044016

\bibitem{Ashtekar2002}
Ashtekar A and Krishnan B 2002
\newblock {\em Phys. Rev. Lett.} {\bf 89(26)} 261101

\bibitem{Ansorg2005a}
Ansorg M and Petroff D 2005
\newblock {\em Phys. Rev. D} {\bf 72} 024019

\bibitem{Arfken1970}
Arfken G 1970
\newblock {\em Mathematical Methods for Physicists}, 2nd ed., Academic Press, Orlando, pp. 115--117

\bibitem{Moon1988}
Moon P and Spencer D E 1988
\newblock {\em Field Theory Handbook: Including Coordinate Systems, Differential Equations, and Their Solutions}, 2nd ed., 
 Springer-Verlag, New York, pp. 110--112

\bibitem{Barrett93}
Barrett R, Berry M, Chan T, Dongarra J, Eijkhout V, Romine C and van~der Vorst H 1993
\newblock {\em Templates for the Solution of Linear Systems: Building
Blocks for Iterative Methods}
{\tt http://www.netlib.org/templates/~}


\end{thebibliography}

\end{document}